# Comprehensive Review of Audio Steganalysis Methods[*]

Hamzeh Ghasemzadeh[1,2]*, Mohammad H. Kayvanrad[3]

[1] Department of Communicative Sciences and Disorders, Michigan State University, MI, USA
[2] Department of Computational Mathematics Science and Engineering, Michigan State University, MI, USA
[3] Department of Biomedical Engineering, Amirkabir University of Technology, Tehran, Iran
*ghasemza@msu.edu

**Abstract:** Recently, merging signal processing techniques with information security services has found a lot of attention. Steganography and steganalysis are among those trends. Like their counterparts in cryptology, steganography and steganalysis are in a constant battle. Steganography methods try to hide the presence of covert messages in innocuous-looking data, whereas steganalysis methods try to break steganography algorithms and reveal the existence of hidden messages. The stream nature of audio signals, their popularity, and their wide spread usage make them very good candidates for steganography. This has led to a very rich literature on both steganography and steganalysis of audio signals. This paper intends to conduct a comprehensive review of audio steganalysis methods aggregated over near fifteen years. To that end, both compressed and con-compressed methods are reviewed, and then their important details are presented in different tables. Furthermore, some of the most recent audio steganalysis methods (both non-compressed and compressed ones) are implemented and comparative analyses on their performances are conducted. Finally, the paper provides some possible directions for future researches on audio steganalysis.

## 1. Introduction

Digital computers have revolutionized every aspect of our lives. Information security is one of the branches of science that has benefited entirely from this invention. In particular, merging signal processing techniques with information security services has found a lot of attention. Some of these new trends include multimedia encryption systems and their cryptanalysis [1, 2], multimedia secret sharing, steganography, steganalysis, and watermarking. Each of these techniques serves a different purpose. For example, multimedia encryption addresses confidentiality and watermarking serves the purpose of copyright protection. Among these techniques, steganography is exceptionally interesting. Arguably, the main purposes of steganography are privacy and preventing traffic analysis. In other words, while encryption prevents unauthorized access to the data, it cannot conceal pattern of communications. Therefore, an adversary who is watching the channel can obtain very valuable information including time of communications, frequency of communications, size of messages, identity of senders and recipients, and much more. Steganography is one possible solution to such problems.

Steganography is best described in terms of subliminal channels. Subliminal channels were first introduced under the prisoner's problem [3]. Two accomplices in a crime are apprehended and imprisoned in two different cells. The warden, who wants to gather some information, allows them to communicate as long as he can read their messages. Apparently, the culprits want to talk about escape plan, but they could not use encrypted messages. Therefore, they use innocuous looking messages and hide their escape plan inside them.

Steganalysis is the countermeasure of the warden against such hidden messages and at the same time it can also help with improving security of existing steganography methods [4]. Because steganography changes content of cover signal, usually it leaves a trail and thus it can be detected. Consequently, steganalysis is a decision problem and if it is solved for a specific method with a high probability, that method is considered broken [5]. If embedding method and statistical model of cover is known, optimal detector can be constructed. But, usually such information in not available and therefore steganalysis systems are usually constructed based on feature extraction and machine learning techniques [6]. Figure 1 presents typical block diagram of such systems.

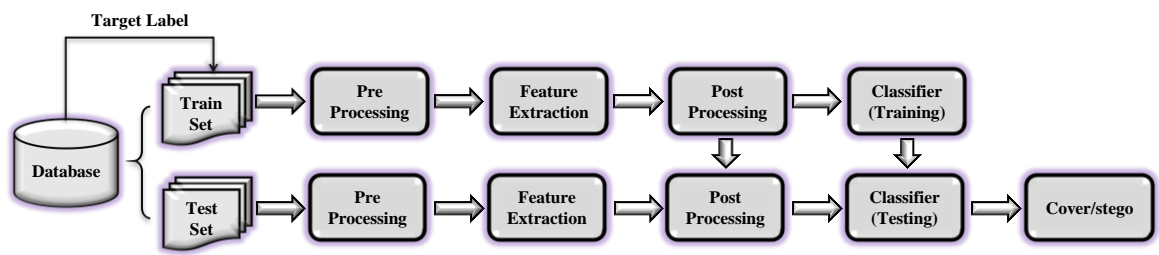

Figure 1. Block diagram of typical steganalysis systems

---

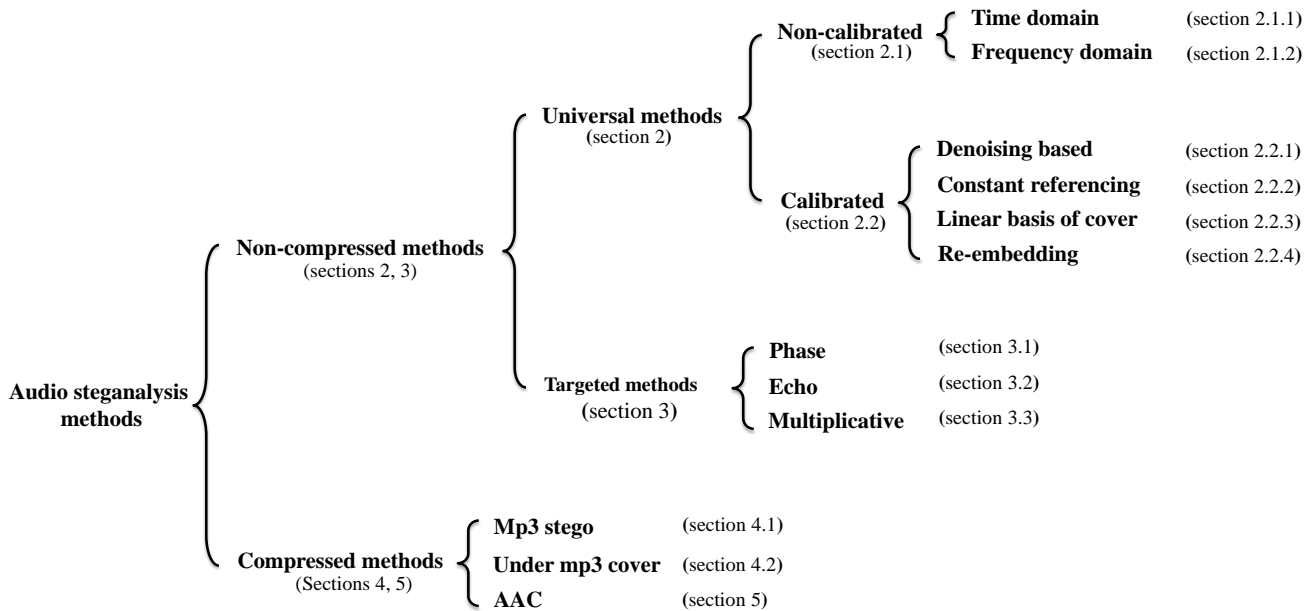

Figure 2. Classification of audio steganalysis methods. The number in the parenthesis shows the relevant sub-section of this paper

Steganalysis is broadly divided into targeted and universal methods. In the targeted paradigm, the warden knows the embedding algorithm. On the other hand, universal methods do not have such assumptions [5]. We can also divide universal methods into blind and semi-blind systems. Blind systems are constructed only from cover signals. On the other hand, semi-blind systems use both classes for determining the decision boundaries. Additionally, two different models of passive and active warden exist. A passive warden listens to communications without interfering with it and his only goal is to detect the presence of hidden messages. But, an active warden tries to prevent communications of hidden messages [5]. Other terms that need to be distinguished are active and passive steganalysis. Passive steganalysis only determines the existence of hidden messages whereas active steganalysis goes one step further and provides the warden with extra information such as the length of hidden message and/or its location.

Current literature lacks a comprehensive review of audio steganalysis methods. This motivated us to review near fifteen years of work on audio steganalysis. This review paper is organized as follows. The next two sections are devoted to the detection of non-compressed steganography methods. To that end, universal and targeted steganalysis methods are reviewed in sections 2 and 3, respectively. Sections 4 and 5 are devoted to steganalysis of compressed methods. In section 6 comparative analyses of audio steganalysis methods are presented. Section 7 describes some possible directions for future research on audio steganalysis and finally the paper is concluded in section 8. Figure 2 summarizes the classification that we have used in this paper with number of relevant sub-sections.

## 2. Universal non-compressed methods

Universal audio steganalysis are broadly divided into non-calibrated and calibrated methods. In the non-calibrated case, steganalysis features are directly extracted from audio signal. On the other hand, in the calibrated case, steganalysis features are extracted by comparing audio signal with its estimated cover/stego signal. Figure 3 compares these two approaches.

### 2.1. Non-calibrated methods

In these methods, features are extracted directly from the signal and they can be categorized according to their feature domain. Although, in some cases such distinction is very hard, yet this methodology is both very common and informative. Figure 3.A depicts a simple schematic of feature extraction in non-calibrated methods.

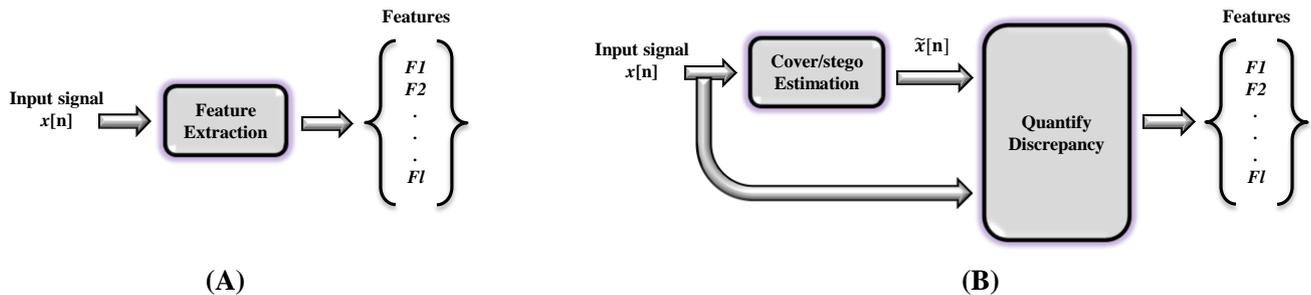

Figure 3. Different approaches to universal audio steganalysis     (A) non-calibrated approach     (B) calibrated approach

### 2.1.1. Time domain features

#### 2.1.1.1. IDS and Steganalysis

Preventing information piracy and its unauthorized disclosure could be an important application of steganalysis. This fact motivated Dittmann to incorporate steganalysis into intrusion detection systems (IDS) [7]. Such applications have real-time requirements; therefore, the system was constructed with simple features and thresholding for making the decision. More specifically, the ratio of ones to zeros in least significance bit (LSB) of signals was used as steganalysis feature.

#### 2.1.1.2. Linear prediction residue

This technique assumed that audio signals are stationary and their samples are highly correlated over short period of time. Embedding process imposes faint changes on the correlation between neighbouring samples. Therefore, if the warden knows the correlation between cover samples, he can detect the abnormalities induced by the embedding process. This idea was implemented in [8]. In this work, linear prediction code (LPC) was used to extract the correlation between neighbouring samples. Furthermore, to extract correlation of different frequency bands, LPC was applied on the wavelet coefficients of all sub-bands. This work used two different feature sets. The first set provided the statistical details of signal in the frequency domain and the second set modelled irregularities in the correlation between samples of signal. These two sets were calculated as higher order statistics (HOS) of wavelet coefficients and LPC prediction errors, respectively. The method was cross-validated with different embedding capacities of Steghide algorithm [9]. Considering the trade-off between missed detection and false alarm, the best result was achieved when the system was trained with stegos embedded at 20% of maximum capacity.

#### 2.1.1.3. Statistical model of histogram

Another possibility is determining the distribution of wavelet coefficients. Fu et al. proposed to extract different statistics from wavelet coefficients and their frequency domain counterparts [10]. This method also benefited from principle component analysis (PCA) for dimension reduction. The results demonstrated that applying PCA reduces dimensionality of the feature space considerably with only negligible degradation in the performance. The method was tested on different wavelet-based steganography algorithms.

#### 2.1.1.4. Chaotic-based Steganalysis

Most of the reviewed works assumed audio signals are stationary and they are outputs of linear systems. However, more recent studies on speech signals have found evidences of some phenomena which linear model cannot describe. Researchers have been able to model such phenomena with chaos theory. Based on such models, noise of steganography would change chaotic structure of audio signal. Therefore, by extracting features from the chaotic structure of audio signals we can distinguish between covers and stegos. This path was followed in [11]. After investigating different chaotic-based measurements, it was concluded that false neighbour fraction and Lyapunov spectrum were more discriminative for steganalysis. The steganalysis system was tested on different settings and it was compared with other methods. The work concluded that performances of all methods were comparable for detection of watermarking, but chaotic-based features detected steganography methods more reliably.

#### 2.1.1.5. Features based on Markov process

Markov processes are the simplest generalization of independent processes and they are very popular for steganalysis. These processes have the interesting property that the past has no influence on the future sample, if the present value is known. If such a system has only finite number of states, it is called a Markov chain. Markov chain of a random variable $X$ is completely determined by its transition probability matrix and it is defined in equation (1).

$$P_{xy} = p(X[n+1] = y \mid X[n] = x) \qquad (1)$$

In [12], Markov transition of the second order derivatives of audio signals were used for steganalysis. To keep the number of features low, transition range of the Markov chain was limited to [-4, 4]. Furthermore, using this range leads to extracting features from smooth regions of signal rather than regions with dramatic changes. This work also argued that besides embedding strength, complexity of audio signals can affect performance of steganalysis. This work proposed the following metric for fast computation of audio signal complexity:

$$C_{x[n]} = \frac{N \sum_{n=0}^{N-2} x''[n]}{(N-2) \sum_{n=0}^{N-1} x[n]} \qquad (2)$$

where, $x''[n]$ and N denote the second order derivative and length of the signal, respectively. Based on this metric, audio signals were divided into three categories of low, middle, and high complexities and each category was tested separately. The simulations showed that these features had good performance even for audio signals with high levels of complexities. This method was later refined with transition range of [-6, 6] and a bigger database for evaluation [13].

#### 2.1.1.6. Autoregressive time delay neural network

Determining the appropriate feature extraction is one of the biggest challenges in steganalysis. Considering the difficulty of modelling temporal characteristics of audio signals, this decision becomes even harder. A special type of networks known as autoregressive time delay neural network (AR-TDNN) was proposed in [14] to address this problem. AR-TDNN has the advantage that feature extraction is not specified explicitly, but the network implements both feature extraction and classification parts of the system. Implementation of these networks has two parts. In the TDNN part, samples of the input are fed into the network and it combines feedback of the output with delayed samples of the input for extracting useful patterns. On the other hand, the autoregressive part of system recognizes the sequence of the previously learned patterns. The efficacy of the system was tested on both LSB and discrete wavelet transform (DWT) embedding algorithms.

### 2.1.2. Frequency domain features

These methods primarily use frequency domain features for distinguishing between covers and stegos.

#### 2.1.2.1. Steganalysis based on MFCC

Mel-frequency cepstrum coefficients (MFCC) are one of the most well-known features in speech processing applications. Cepstrums are frequency components of the logarithm of the magnitude of spectral power of the signal and it can be interpreted as the rate of power changes in different frequency regions. MFCC is a modified version of cepstrum and it reflects some of characteristics of the human auditory system (HAS). Let $X[jw]$ denotes frequency components of signal $x[n]$, equations (3, 4) shows calculation of MFCC.

$$E_k = \sum_{j=0}^{N-1}(|X[jw]|.W_k[j]), 1 \leq k \leq M \quad (3)$$

$$C_k = |F^{-1}(log(E_k))|, 1 \leq k \leq M \quad (4)$$

where $N$ is the length of $x[n]$, $\{W_k\}$ are a set of triangular weighting functions that their centres have equal distances in the Mel scale (figure 4.B), $M$ is the number of weighting functions, , $F^{-1}$ denotes inverse Fourier transform, and $C_k$s are MFCCs.

Based on potency of MFCC for speech recognition applications, it was proposed for audio steganalysis in [15]. This work focused on steganalysis of VoIP channels and it used three sets of features. The first set was statistical characteristics of the signal, and it included entropy, LSB ratio, flipping rate of LSB, and some other time domain statistical moments. The second set consisted of 29 MFCCs and the third set was based on the hypothesis that removing speech-relevant portion of the signal improves steganalysis. Therefore, the signal in the range of 200 to 6819.59 hertz was filtered. Then MFCCs of the filtered signal were calculated. Efficacy of this method was tested on five steganography and four watermarking methods. Furthermore, different combinations of feature sets were investigated. The tests confirmed that removing speech-relevant portions of signals improved performance of steganalysis.

Another work argued that derivative of the signal is more informative for steganalysis [16]. This work showed that taking the second order derivative of signal improves discriminative properties of high-frequency regions. An experiment was conducted to justify this idea where, the whole spectrum was divided into 80 different regions. Comparing discriminative property of different regions showed that high frequency regions were more informative. To investigate the potency of the method, the system was contrasted with ordinary MFCC and wavelet-based MFCC.

Results showed that the derivative-based MFCC had the best results.

To implement a more powerful system different features were combined in [17]. First, short-term characteristics of signal were extracted from 12 MFCCs. The second set consisted different moments of spectral characteristics from the second order derivative of the signal. Finally, features based on audio quality metric [18] and LPC residue [8] were extracted. This work fused all of these features and formed a vector of 52 features. Then, feature selection was conducted based on F-scores and the system was tested in both targeted and universal scenarios.

#### 2.1.2.2. Reversed-psychoacoustic model of human hearing

Ghasemzadeh et al. argued that employing features based on models of human auditory system (HAS) is counter intuitive for steganalysis and it would lead to discarding vital information [19]. According to this work, a steganography method is insecure if its stegos are distinguishable from its covers. Therefore, in its most basic form, the human perception should be oblivious to steganography induced noise. Based on this, it was argued that if features are extracted based on models of HAS their significance would diminish. To address this, an artificial auditory system was designed that could virtually hear effect of steganography and therefore could discriminate stegos from covers. This new model was called reversed Mel and it had the maximum deviation from HAS. That is, it had finer resolution in high frequencies and coarser resolution in low frequencies. Equations (5, 6) compares Mel and R-Mel scales.

$$Mel = 1127 \times ln\left(1 + \frac{f}{700}\right) \quad (5)$$

$$RMel = 1127 \times ln\left(1 + \frac{F_s/2 - f}{700}\right) \quad (6)$$

where, $f$ and $F_S$ denote the given frequency in hertz and the sampling frequency of the signal, respectively. Figure 4.A presents a comparison between Mel and R-Mel scales. In addition, calculation of cepstrum coefficients relies on a set of triangular windows. Figure 4.B compares windows constructed based on Mel and R-Mel scales.

This work was further improved in [20]. Let $c[n]$, $s[n]$, and $e[n]$ denote cover, stego, and its steganography noise, respectively. A common formulation of steganography is:

$$s[n] = c[n] + e[n] \quad (7)$$

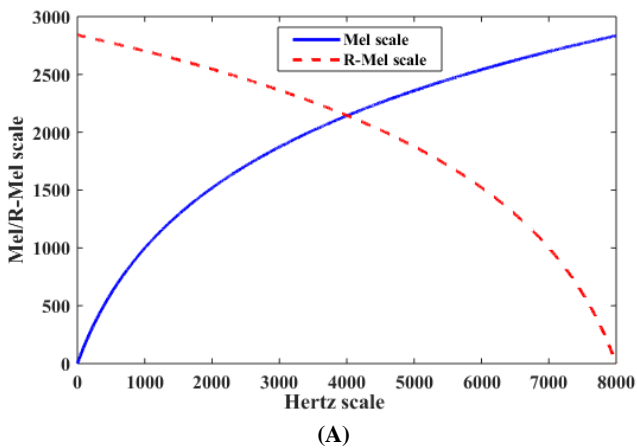
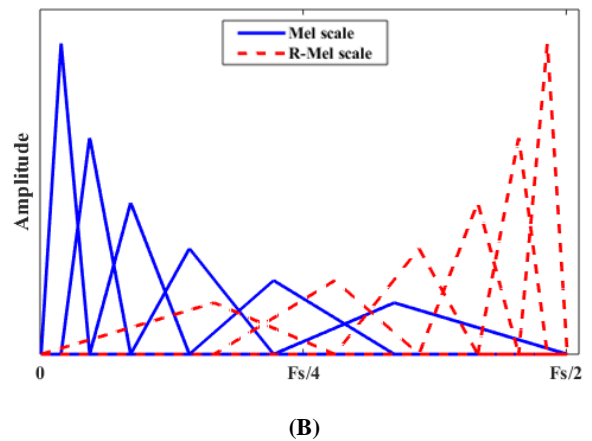

(A)  (B)

Figure 4. Comparison between R-Mel and Mel    (A) R-Mel scale vs Mel scale    (B) Filter banks constructed based on R-Mel and Mel

To justify that maximum deviation from HAS is beneficial for audio steganalysis, spectrums of covers and their steganography noise was divided into $L$ disjoint sub-bands [20]:

$$(i-1) \times \frac{\pi}{L} \leq B_i \leq i \times \frac{\pi}{L}, 1 \leq i \leq L \quad (8)$$

Then, sub-band signal to noise ratio ($SNR_i$) was defined as:

$$SNR_i = 10 log_{10}\left(\frac{\int_{B_i}|C(jw)|^2}{\int_{B_i}|E(jw)|^2}\right), \quad 1 \leq i \leq L \quad (9)$$

Where, $C(jw)$ and $E(jw)$ denote the spectrum of cover and its steganography noise. This work compared plot of $SNR_i$ for different steganography methods with frequency resolution of Mel and R-Mel scales and showed that R-Mel scale was more suitable for steganalysis purposes. The work also used HOS for providing better discriminative properties. Finally, genetic algorithm (GA) [21] was invoked to find the optimum sub-set of features. This method was tested on a wide range of data hiding algorithms including both LSB and non-LSB methods. Furthermore, results of both targeted and universal scenarios were investigated. The results showed that HOS and feature selection based on GA improve performance of steganalysis considerably.

## 2.2. Calibrated methods

Finding features that are independent from contents of signal and only reflects the presence of hidden message is one of the most challenging problems in steganalysis. Apparently, if the cover is known such features can be extracted very easily. But, in the realistic scenarios this assumption does not hold. Therefore, researchers have proposed estimation as a practical solution to this dilemma.

### 2.2.1. Self-generated estimation of cover

This idea was proposed in [18] and it is based on representing steganography as an additive noise (equation 7). If that noise is found effectively, virtually a good estimation of the cover signal would be available. This is the main rationale behind steganalysis methods of this category. Basically, these methods estimate the cover by applying a noise removal procedure on the signal. Also, it is expected that different amount of noise is extracted from stegos and covers. These methods have employed different approaches to quantify such discrepancies.

#### 2.2.1.1. Steganalysis based on Audio Quality Metrics (AQM)

This idea was proposed by Ozer el al. [18] and the wavelet-thresholding was employed for noise removal and estimation of the cover. If discrepancies between estimated covers and stego are measured accurately, they can be distinguished. This work noticed that, in speech coding applications, researchers have been trying to address a similar problem for measuring artefacts of speech coders for a long time. Result of those endeavours is a set of measurements known as audio quality metrics (AQM), which can be categorized into three groups of time, frequency, and perceptual domains. AQMs measure discrepancies between short frames of the reference signal and its modified version and the final metrics are calculated as the average value of each metric over all frames.

The same methodology was adapted for steganalysis where signal under inspection and its denoised version were considered as the reference and modified signals. Then, their discrepancies were evaluated through AQMs. Finally, these metrics were used as steganalysis features. Work of [18] tested different combinations of feature selections and machine learning methods.

#### 2.2.1.2. Steganalysis based on Hausdorff Distance

Liu et al. argued that AQMs have been designed specifically for objective assessment of the quality of audio signals and not steganography impairments [22]. They noted that, this argument is especially true for perceptual metrics. Therefore, it is very likely that AQM features have limited capability to capture the effects of embedding. To alleviate such problems, other researchers have used Hausdorff distance for measuring the effect of message embedding [22, 23]. Hausdorff distance is fundamentally a max-min measure which has found numerous applications in template matching and content-based retrieval problems. Given two finite sets $A = \{a_1, ..., a_m\}$ and $B = \{b_1, ..., b_n\}$ the symmetric Hausdorff distance between them is defined as:

$$H(A, B) = ma x(h(A, B), h(B, A)) \quad (10)$$

where,

$$h(A, B) = \max_{a_i \in A} \min_{b_j \in B} \|a_i - b_j\| \quad (11)$$

These works also used noise removal for estimating the cover. Then, input signal and its estimated cover were segmented and decomposed with wavelet. Then, Hausdorff distance between each pair of sub-bands was calculated. Final features were HOS of Hausdorff distances over all frames. The system was tested on Steghide [9] and the result was compared with other steganalysis methods. Furthermore, discriminative capability of different sub-bands were investigated and it was shown that features extracted from lower levels of wavelet decomposition were more significant.

Geetha et al. used the same idea for feature extraction, but investigated efficacy of six decision tree classifiers on different steganography and watermarking methods [23].

#### 2.2.1.3. Modelling the noise based on GMM and GGD

Another solution for distinguishing between stego and cover is to compare distributions of their wavelet coefficients directly. Work of [24] used Gaussian mixture model (GMM) and generalized Gaussian distribution (GGD) for this purpose. To that end, two different methods were proposed. The first method modelled distribution of wavelet coefficients of all sub-bands with GMM and used them for steganalysis. In the second method, GGD and GMM were employed for capturing distributions of wavelet coefficients of signals and their de-noised counterparts. Finally, deviations between the two distributions were quantified with four different distance measures. Unfortunately, this work did not provide any result on the performance of the system.

The same idea was investigated more properly in [25]. The authors showed that spread spectrum embedding causes the histogram of wavelet coefficients to become flatter around zero. GMM and GGD were employed to capture this flatness. The results showed that GMM captures artefacts of spread spectrum hiding more accurately, and the best performance was achieved when three Gaussian kernels were mixed. This work measured potency of individual features on two different spread spectrum methods, but the system was not tested on combinations of more than one feature.

### 2.2.2. Constant referencing (CIAQM)

Previous methods used original signal for estimating the cover; and therefore they are called self-referencing. Avcibas showed that self-referencing (e.g., denoising) leads to dependence of features on the content of signals [26]. To alleviate this problem, he proposed the method of constant referencing. That is, two fixed reference signals were selected, one of which was a cover and the other one was its stego version. Let $M(x, y)$ denotes a function that measures discrepancies between its two input signals $x$ and $y$ in a meaningful manner. Also, let $r$ and $r+e$, denote the reference cover and its stego version, respectively. For every incoming signal ($s$), the proposed features ($f$) were defined as:
$$f(s, r, r+e) = M(s, r) - M(s, r+e) \tag{12}$$

Work of [26] showed that when mean and standard deviation are used for discrepancy measure ($M$), equation (12) is independent from the cover signal, but for the actual steganalysis system, AQMs were used to measure discrepancies ($M$) and classification was achieved through linear regression. The efficacy of this method was tested on six different embedding algorithms covering both steganography and watermarking methods. Simulations showed that constant referencing outperformed self-referencing methods. Also, results showed that this improvement was more noticeable for steganography methods.

### 2.2.3. Estimation of cover space model

While previous works measured discrepancies between a reference and the signal for making the decision, another approach is possible if the high dimensional model of cover space is known. In this fashion we can check every signal $x[n]$ with this model, if it fits the model, it is a cover and otherwise it is a stego. Unfortunately, a perfect model does not exist for empirical cover sources [6]. But this idea could be exploited for estimating a model for empirical cover signals. This approach was pursued by Johnson et al [27]. First, short time Fourier transform (STFT) was invoked to capture time-frequency regularities of audio signals. Then, PCA was used to extract a set of linear bases for this time-frequency vector space. These linear bases formed a vector quantizer and the residual signal from this quantization was used to quantify accuracy of this sub-space for modelling the audio signal. Therefore, features were calculated as HOS of quantization error. Simulations were conducted on LSB embedding methods and results showed that this method achieves reasonable accuracy in LSB embedding when at least 4-bits are used for data hiding process.

### 2.2.4. Re-embedding calibration

Previous calibration methods are similar in the sense that estimation of covers were used for calibration. Calibrating features based on characteristics of stegos is another possibility that was pursued in [28]. That paper addressed both targeted and universal cases. In the targeted paradigm the embedding algorithm is known; therefore, signals were embedded with the same algorithm and a random message. Then, difference between features extracted from original signal and its embedded version were used for steganalysis. Let $x[n]$, $\mathcal{A}_{em}$ and $m[n]$ denote a signal, the embedding algorithm, and a random message, then steganalysis features were defined as:
$$\tilde{x}[n] = \mathcal{A}_{em}(x[n], m[n]) \tag{13}$$
$$F = \mathfrak{F}(x[n]) - \mathfrak{F}(\tilde{x}[n]) \tag{14}$$
where, $\mathfrak{F}$ denotes a suitable feature extraction process. This technique was later extended into universal paradigm, where embedding algorithm is not known. For this purpose, the concept of bit-plane sensitivity was defined as the amount of noise that is introduced into each bit-plane after random embedding. Let $\mathcal{B}_i(c)$, $\mathcal{B}_i(s)$, and BER(,) denotes bit-plane $i$ of cover, bit-plane $i$ of stego, and bit error rate. Equation (15) shows definition of bit-plane sensitivity for bit-plane $i$:
$$\mathbb{S}_i = 2 \times BER(\mathcal{B}_i(c), \mathcal{B}_i(s)) \tag{15}$$

That work investigated sensitivity of different bit-planes of a wide range of data hiding algorithms and showed that 1-LSB was the most sensitive bit-plane. Consequently, LSB re-embedding was employed as a universal calibration method. Additionally, potency of cepstrum features were compared with energy of filter banks and it was shown that energy of filter banks were superior. Finally, due to positive effect of feature normalization and good performance of GA for feature selection [29], features were normalized and GA was invoked for feature selection.

## 2.3. Summary

In this section a summary of the investigated methods is presented. First, we take a brief look on the relationship between different methods. Figure 5 presents how different works are inter-related to each other.

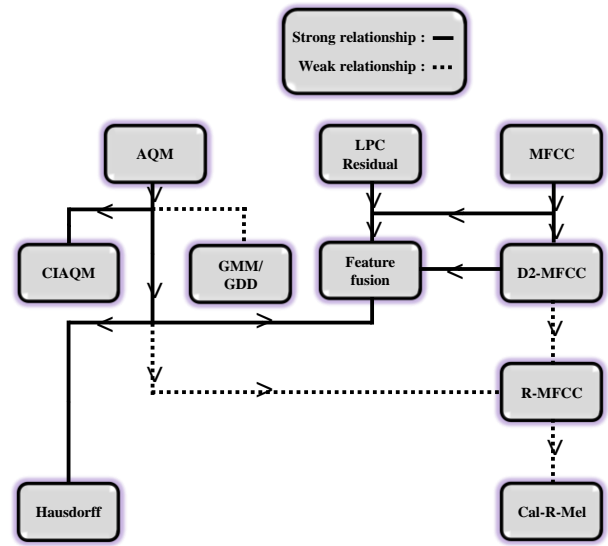

Figure 5. Relationship between different papers

| Method | Ref. | Section | Method | Ref. | Section |
|--------|------|---------|--------|------|---------|
| LPC | [8] | 2.1.1.2 | AQM | [18] | 2.2.1.1 |
| MFCC | [15] | 2.1.2.1 | Hausdorff | [22, 23] | 2.2.1.2 |
| D2-MFCC | [16] | 2.1.2.1 | GMM | [24, 25] | 2.2.1.3 |
| Fusion | [17] | 2.1.2.1 | CIAQM | [26] | 2.2.2 |
| R-MFCC | [19, 20] | 2.1.2.2 | Cal-R-Mel | [28] | 2.2.4 |

One of the major differences between the reviewed papers was the data hiding algorithms that they were tested on. Investigated data hiding algorithms are presented in three different tables. Table 1 summarizes the watermarking methods, the symbol that we have used for referring to them, and their embedding domain. These methods include: 2A2W [30], COX [31], DSSS [32], Echo [32], FHSS, LSB watermarking [30], spread spectrum audio watermarking (SSAW) [33], WAWW [30], and multi carrier spread spectrum (MCSS) [34].

Table 1. Specifications of watermarking methods

| Symbol | Name | Domain | Ref. |
|---|---|---|---|
| W1 | 2A2W | wavelet | [30] |
| W2 | COX | Freq. | [31] |
| W3 | DSSS | Time | [32] |
| W4 | Echo | Time | [32] |
| W5 | FHSS | Freq. | - |
| W6 | LSB | 1-LSB | [30] |
| W7 | SSAW | Freq. | [33] |
| W8 | VAWW | wavelet | [30] |
| W9 | MCSS | Freq. | [34] |

Table 2 summarizes the LSB steganography methods, the symbol that we have used for referring to them, and their embedding domain. These methods include: Heutling [30], Hide4pgp [35], Invisible Secret [36], LSB Matching [37], Steganos [38], Steghide [9], Stools [39].

Table 2. Specifications of LSB-based steganography methods

| Symbol | Name | Domain | Ref. |
|---|---|---|---|
| SL1 | Heutling | 1-LSB | [30] |
| SL2 | Hide4pgp | 1- 4LSB | [35] |
| SL3 | Invisible Secret | 1-LSB | [36] |
| SL4 | LSB Matching | 1-LSB | [37] |
| SL5 | Steganos | LSB | [38] |
| SL6 | Steghide | 1-LSB | [9] |
| SL7 | Stools | LSB | [39] |

Information of non-LSB methods are presented in Table 3. These methods include: Addition Method (AM) [10], Amplitude Modulation (AMod) [40], DWT-FFT [41], DWT fusion [42], Mp3Stego [43], Publimark [30], Quantization Index Method (QIM) [10], Stego wave [44], Stochastic Modulation (StMod) [45], WaSpStego [15], Wavelet LSB [10], Integer wavelet (I-Wavelet) [46], and DSSS in the frequency domain (DSSS + DCT) [47].

Table 3. Specifications of non LSB-based steganography methods

| Symbol | Name | Domain | Ref. |
|---|---|---|---|
| SN1 | AM | wavelet | [10] |
| SN2 | AMod | Freq. | [40] |
| SN3 | DWT-FFT | Wavelet + Freq. | [41] |
| SN4 | DWT fusion | wavelet | [42] |
| SN5 | Mp3stego | Side Info | [43] |
| SN6 | Publimark | no information | [30] |
| SN7 | QIM | wavelet | [10] |
| SN8 | Stego wave | no information | [44] |
| SN9 | StMod | Time | [45] |
| SN10 | WaSpStego | wavelet | [15] |
| SN11 | Wavelet LSB | wavelet | [10] |
| SN12 | I-Wavelet | DWT | [46] |
| SN13 | DSSS + DCT | Freq. | [47] |

Table 4 presents the embedding algorithms that each paper has investigated.

Different criteria may be defined to report efficacy of classification. Definition of these criteria are as follows.
- True negative (TN): the number of cover samples that are classified as cover samples.
- True positive (TP): the number of stego samples that are classified as stego samples.
- False negative (FN): the number of stego samples that are classified as cover samples.
- False positive (FP): the number of cover samples that are classified as stego samples.

True positive rate (TPR) is the probability of detection of stego samples correctly and it is calculated as:

$$TPR = \frac{TP}{TP + FN} \times 100\% \qquad (16)$$

True negative rate (TNR) is the probability of detection of cover samples correctly and it is calculated as:

$$TNR = \frac{TN}{TN + FP} \times 100\% \qquad (17)$$

Accuracy (Ac.) is the probability of correct classification and it is calculated as:

$$Ac. = \frac{TP + TN}{TP + FN + TN + FP} \times 100\% \qquad (18)$$

Other important differences between the reviewed methods were evaluation criteria, domain of feature extraction, number of features, type of classifier, number of clean files in their databases, size of training set, and their performances. These parameters are summarized in table 5.

## 3. Targeted non-compressed methods

HAS has a low differential sensitivity and it can only perceive relative phase of the signal [32]. Based on these characteristics different embedding methods have been proposed. This section presents steganalysis systems that have been designed specifically for their detection.

### 3.1. Phase coding methods

Steganalysis of phase coding system was investigated in [48]. This work observed that phase embedding preserves the relative phase of each block, but the phase difference between consecutive blocks are changed. To capture signatures of phase embedding, signal was segmented and unwrapped phase of all segments were calculated. The steganalysis features were statistical moments of the absolute difference between phases of consecutive segments. The system was tested for different length of blocks and subblocks.

### 3.2. Echo embedding methods

Echo hiding methods have a bank of kernels and depending on the bit of message, one of them is selected and is convolved with different segments of cover [4]. Therefore, kernels play an important role on the characteristics of echo hiding methods. For example, it is possible to design kernels that achieve better robustness. Positive-negative (PN) and forward-backward (FB) are examples of kernels that have been designed for such purposes. Let $\delta[n]$ denotes the impulse signal, then $\alpha\delta[n - d_i]$ denotes an impulse signal which is centred at time $d_i$ and has amplitude of $\alpha$. Impulse responses of famous kernels are as follows:

$$h_{Basic}[n] = \delta[n] + \alpha\delta[n - d_i] \qquad (19)$$
$$h_{PN}[n] = \delta[n] + \alpha\delta[n - d_i] - \alpha\delta[n - d'_i] \qquad (20)$$
$$h_{FB}[n] = \delta[n] + \alpha\delta[n - d_i] + \alpha\delta[n + d_i] \qquad (21)$$

Table 4. Details of steganography methods investigated by each steganalysis paper

| | Steganalysis methods | | | | | | | | | | | | | | | | | |
|---|---|---|---|---|---|---|---|---|---|---|---|---|---|---|---|---|---|---|
| | [7] | [8] | [10] | [11] | [12] | [13] | [14] | [15] | [16] | [17] | [18] | [19] | [20] | [22] | [23] | [25] | [26] | [27] | [28] |
| W1 | | | | | | | | ✓ | | | | | | | | | | | |
| W2 | | | ✓ | | | | | | | ✓ | ✓ | ✓ | | ✓ | | | | | ✓ |
| W3 | | | ✓ | | | | ✓ | | ✓ | ✓ | | | | | ✓ | ✓ | | | |
| W4 | | | ✓ | | | | | | ✓ | ✓ | | | | ✓ | | ✓ | | | |
| W5 | | | ✓ | | | | | | | ✓ | | | | | | ✓ | ✓ | | |
| W6 | | | | | | | | ✓ | | ✓ | | | | | | | | | |
| W7 | | | | | | ✓ | | | | | | ✓ | ✓ | | | | | ✓ | |
| W8 | | | | | | | | ✓ | | | | | | | | | | | |
| W9 | | | | | | | | | | | | | | | | | | | ✓ |
| SL1 | | | | | | | | ✓ | | | | | | | | | | | |
| SL2 | | ✓ | | ✓ | ✓ | ✓ | ✓ | | ✓ | | ✓ | ✓ | ✓ | | | ✓ | ✓ | ✓ | |
| SL3 | | | | | ✓ | ✓ | | | ✓ | | | | | | | | | | |
| SL4 | | | | | ✓ | ✓ | | | ✓ | | | | | | | | | | |
| SL5 | | | | ✓ | | | | | | ✓ | | | | | | ✓ | | | |
| SL6 | ✓ | ✓ | | ✓ | ✓ | ✓ | ✓ | ✓ | ✓ | | ✓ | ✓ | ✓ | ✓ | ✓ | | ✓ | | ✓ |
| SL7 | | ✓ | | | | | | | | | ✓ | | | ✓ | | | | | |
| SN1 | | | ✓ | | | | | | | | | | | | | | | | |
| SN2 | | | | | | | | | | | | | | | ✓ | | | | |
| SN3 | | | | | | | ✓ | | | | | | | | | | | | |
| SN4 | | | | | | | | | | | | | | | ✓ | | | | |
| SN5 | ✓ | | | ✓ | | | | | | | | | | | | | | | |
| SN6 | | | | | | | | | ✓ | | | | | | | | | | |
| SN7 | | | ✓ | | | | | | | | | ✓ | | | | | | | |
| SN8 | | | ✓ | | | | | | | | | | | | | | | | |
| SN9 | | | | | | ✓ | | | | | | | | | | | | | |
| SN10 | | | | | | | | | ✓ | | | | | | | | | | |
| SN11 | | | ✓ | | | | | | | | | | | | | | | | |
| SN12 | | | | | | | | | | | | | | | ✓ | | | | ✓ |
| SN13 | | | | | | | | | | | | | | | ✓ | | | | ✓ |

Table 5. Summary of universal steganalysis papers. Notations of the table are as follows -: not reported parameter, M: music samples, S: speech samples. Performance criteria are the average value over all embedding methods and capacities.

| Method | Reference | Feature Domain | Feature No. | Classifier | Clean No. | Train Size | Universal Results | | Targeted Results | |
|---|---|---|---|---|---|---|---|---|---|---|
| | | | | | | | TPR | TNR | TPR | TNR |
| IDS | [7] | Bits | 3 | Threshold | 30 | - | - | - | - | - |
| LPC | [8] | Wavelet | 40 | SVM | 500 | 50% | - | - | 99.24 | 94.91 |
| Wavelet + PCA | [10] | Wavelet | 36 | NN | 400 | 62.5% | - | - | 96.22 | 95.11 |
| Chaotic | [11] | chaotic | 22 | SVM | 2554 | 50% | - | - | M:69.14 S: 91.37 | M:58.56 S: 89.89 |
| Markov | [12] | Time | 81 | SVM | 12000 | 50% | - | - | Ac. = 92.2 | |
| Markov | [13] | Time | 169 | SVM | 19380 | 70% | - | - | Ac. = 97.3 | |
| AR-TDNN | [14] | Time | - | NN | 150 | - | - | - | 68.48 | 78.29 |
| MFCC | [15] | Cepstrum | 36 | SVM | 389 | 80% | - | - | 66.04 | - |
| D2-MFCC | [16] | Freq. | 29 | SVM | 12000 | 50% | - | - | Ac. = 85.9 | |
| Feature Fusion | [17] | Freq. | 12 | SVM | 600 | 67% | Ac. = 91 | | Ac. = 90 | |
| AQM | [18] | Time + Freq. | 19 | SVM | 664 | 50% | 81.8 | 79.7 | 93.5 | 92.75 |
| RMFCC | [19] | Freq. | 29 | SVM | 4169 | 70% | - | - | 97.29 | 94.71 |
| RMFCC + HOS+GA | [20] | Freq. | 21 | SVM | 4169 | 70% | 94.4 | 99.1 | 99.1 | 99.0 |
| Hausdorff | [22] | Wavelet | 25 | SVM | 994 | 90% | - | - | 95 | 88 |
| Hausdorff + tree | [23] | Wavelet | 25 | J48 Tree | 200 | 75% | - | - | 88.64 | 72.89 |
| DWT+GMM | [25] | Wavelet | 1 | SVM | 1000 | 50% | - | - | 93.44 | 91.22 |
| CIAQM | [26] | Time + Freq. | 19 | linear regression | 100 | 50% | - | - | 95 | 95 |
| STFT+PCA | [27] | STFT | 4 | SVM | 1800 | 80% | - | - | 56.95 | 98.1% |
| Cal-R-Mel | [28] | Freq. | 15 | SVM | 4169 | 90% | M:98.7 S:87.8 | M:99.8 S: 96.7 | M:99.5 S:94.9 | M:99.7 S: 93.9 |

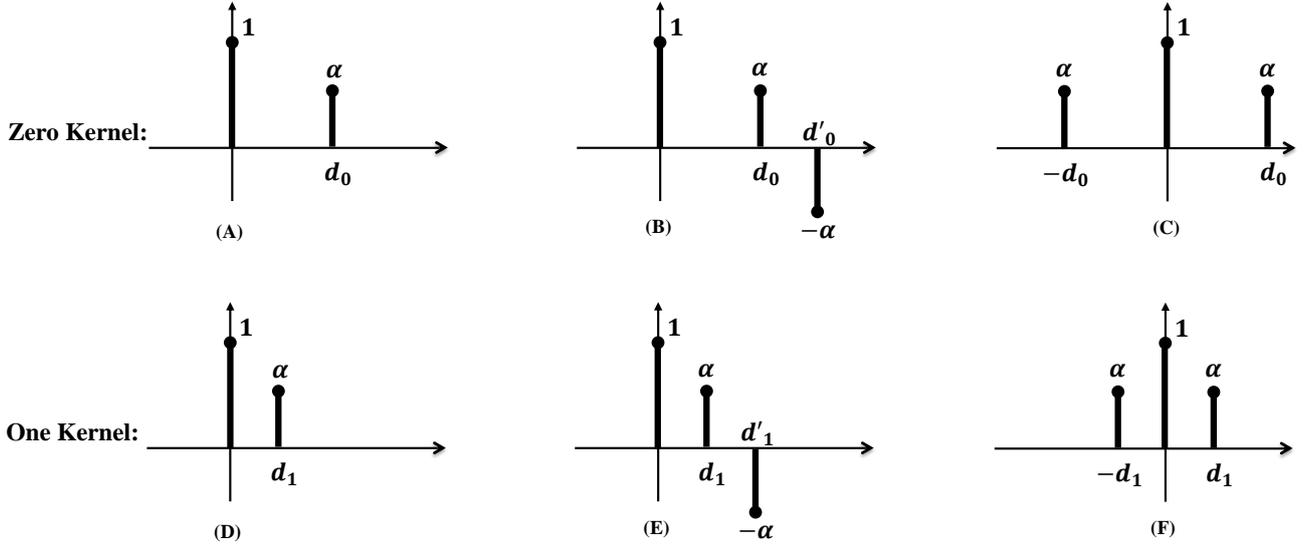

Figure 6. Impulse response of different echo kernels  (A, D) basic kernel  (B, E) positive-negative kernel  (C, F) forward-backward kernel

If message is binary encoded, then $i \in \{0,1\}$. Figure 6 compares these different kernels.

Targeted steganalysis of the basic kernel was first investigated in [49]. It was shown that for short windows, stegos have a peak in their power cepstrums. Therefore, cepstrum of the signal was calculated on short windows and different moments of local maxima of all windowed-cepstrums were used for steganalysis. The method investigated effect of different parameters of the system.

Steganalysis of echo hiding with PN kernels was discussed in [50]. They showed that the power cepstrum and the complex cepstrum of the stego signal exhibit peaks in delay positions of the kernels ($d_i$). To detect these artefacts, the audio signal was segmented into 20ms chunks. Skewness of both the power cepstrum and the absolute value of the complex cepstrum over all frames were calculated. The steganalysis feature was defined as kurtosis of the calculated values of skewness. Simulation results showed that feature based on the absolute value of complex cepstrum was more discriminative than power cepstrum-based feature.

Xie et al. went one step further and presented an active steganalysis on echo data hiding [51]. The proposed system was based on analysing the behaviour of cepstrum of a short sliding window. Based on that work, using a sliding window ($W_s$) smaller than the segment length, four situations will exist: 1) when $W_s$ is inside a zero embedded segment, 2) when it is inside a one embedded segment, 3) when it is crossing from a one (zero) embedded into another one (zero) embedded segment, and 4) when it is crossing the border of two segments that have been embedded with different bits. The work showed that if histogram of peak position of all sliding cepstrums is plotted, it exhibits high densities around the value of delays. Furthermore, cepstrum peak location aggregation rate (CPLAR) was introduced as the ratio of the number of times that location of cepstrum peak was at one of those positions to the total number of windowed cepstrums. CPLAR was used for deciding about the presence/absence of the hidden message. The method was also able to find the length of segmentation in the embedding system. Extensive simulations were done on different echo kernels and the results showed that detection and estimation of parameters are harder when PN kernel is used.

Recently, a more secure echo method was proposed in [4]. This work proposed that segment length and parameters of echo kernels be varied in a pseudo random fashion. Simulations of this work showed that this strategy makes CPLAR feature obsolete. Also, they showed this technique also improves robustness of the system against active warden.

### 3.3. Multiplicative embedding methods

A multiplicative steganography scheme can be expressed as:
$$s[n] = c[n](1 + m[n]) \qquad (22)$$
where $s[n]$, $c[n]$ and $m[n]$ denote stego, cover, and the hidden message. According to equation (7) the steganography induced noise is equal to:
$$e[n] = c[n].m[n] \qquad (23)$$
Apparently, this noise is multiplicative and cover dependent. Steganalysis of such embedding methods was presented in [52]. It was claimed that the conventional steganalysis methods cannot detect multiplicative systems properly. To address this problem, the logarithm of absolute value of audio samples was used. In this fashion, multiplicative noise was transformed into additive noise and then, wavelet was applied on this new signal. Statistical moments of signal and its sub-band coefficients were extracted and they were appended with statistics of the linear prediction residue of each sub-band [8]. The experimental results showed this technique improves detection of multiplicative steganography systems.

### 3.4. Summary

Targeted steganalysis papers are summarized in table 6.

### 4. Mp3 Steganalysis

The mpeg-1 layer III (mp3) format provides high compression rate and good quality. These characteristics have turned mp3 into one of the most popular formats of audio signals. Consequently, it is a suitable medium for steganography. Over the past decade different mp3 steganography methods have been developed. They include Mp3stego [43], under mp3 cover (Ump3c) [53], Mp3stegz

Table 6. Summary of targetd steganalysis papers. Notations of the table are as follows -: not reported parameter

| Ref. | Target method | Domain | Feature No. | Classifier | Cover No | Train size | Performance | |
|---|---|---|---|---|---|---|---|---|
| | | | | | | | TPR | TNR |
| [48] | Phase | Freq. | 5 | SVM | 800 | 25% | 98.2 | 95 |
| [49] | Echo | Freq. | 8 | SVM | 1200 | 50% | 82.17 | 87.83 |
| [50] | Echo | Freq. | 1 | SVM | 300 | 17% | 94 | 89.2 |
| [51] | Echo | Freq. | 1 | Threshold | 200 | - | - | - |
| [52] | Multiplicative | wavelet | 40 | SVM | 450 | 44% | 94.8 | 96.4 |

[54], Huffman table swapping [55], quantization step parity [56], Linbit [57], Bv-Stego [58], and other methods [59-61]. Current literature has only investigated steganalysis of Mp3stego and Ump3c. This section reviews their findings. Also, italic font is used to distinguish between the terms adopted from mp3 standard and the rest of the text.

### 4.1. Steganalysis of Mp3stego

Mp3 encoding procedure consists of two nested loops. In this fashion, the *inner loop* does the actual quantization of the data and determines the suitable quantizer step that meets with the available bit budget. On the other hand, the *outer loop* controls distortion of the encoding process and keeps it beyond the level of perception. Mp3stego changes the termination condition of *inner loop* and hides the message during the compression process.

Using notation of mp3 standard, the *part2_3_length* is the field that determines the number of bits that are used for encoding of each frame. Mp3stego embeds bits of message as the parity of the *part2_3_length* variable by controlling when the *inner loop* is terminated [43]. Because Mp3stego hides the information during the compression process, many parts of mp3 are changed. Consequently, steganalysis of Mp3stego can be accomplished in different ways. They include variance of *block size*s, numbers of different block lengths, MDCT coefficients, statistics of the *bit reservoir*, and statistics of *quantization steps*.

#### 4.1.1. Histogram of block length

Westfeld investigated Mp3stego algorithm and concluded that block length of modified frames are smaller than their normal counterparts [62]. In other words, if the maximum length of frames were fixed, stego files would have lower bit rates. Of course this is not the case with mp3 standard and the algorithm adjusts length of blocks to achieve the target average bit rate (e.g. 128k). Therefore, if bit rate of one block is decreased, next blocks will use those extra bits. This work showed that variance of *block size* is different between covers and stegos.

#### 4.1.2. Number of different block lengths

Investigating histogram of stegos and covers reveals that their block lengths are different. Specifically, Mp3stego leads to more different block lengths to be used. Based on this idea an ultra-light-weight steganalysis system was proposed in [7]. This system used length of the first block (F) as an estimation of the expected value of block length. Then the number of different block lengths in the file was calculated (C). After investigating the ratio of F/C for covers and stegos, they proposed the empirical value of 4.6 as the appropriate threshold. In this fashion, the ratio of F/C was larger than 4.6 for covers.

#### 4.1.3. Statistics of MDCT coefficients

Mp3stego increases *quantization step* of encoder. Apparently, this decreases the absolute value of MDCT coefficients. If statistical distributions of the MDCT coefficients are known, this extra distortion could be detected. This idea was pursued by Qiao et al. [63]. They extracted different statistical metrics from signal to capture any anomaly in its MDCT coefficients. First, GGD was used to model distribution of the MDCT coefficients in each frame. Second set of features were calculated from different subbands of the second order derivatives of the MDCT coefficients. Finally, the third and fourth sets of features were Markov transition probabilities of inter-frame and intra-frame MDCT coefficients. In the simulations, different combinations of features were considered and the best result was achieved when all features were used. This work was improved in [64], which added feature selection to the method and used a bigger database for evaluation.

Jin et al. used the same concept and extracted joint probability of adjacent MDCT coefficients in the same channel and the neighbour channel [65]. The work showed that:

1) Features along the minor diagonal direction were more discriminative.

2) Features in the center of the transitions matrix were more powerful.

Based on these observations a method for reducing the number of features was proposed. This work was extended in [66] and features were extracted from difference of absolute values of MDCT in the inter and intra frames of mp3.

#### 4.1.4. Calibrated Mp3stego steganalysis

Audio signals have a wide dynamic range, implying that some of their portions have fast transitions and are more complex. Apparently, reconstruction of these portions needs more bits. To solve this problem, mp3 standard benefits from a short buffer mechanism called *bit reservoir*. This technique keeps the average bit rate constant, but it allows individual frames to have different bit rates. In this fashion, frames with low complexities are encoded with fewer bits and the extra bits are saved for more complex frames.

Frames that are embedded with Mp3stego typically have larger *quantization steps*. Therefore, their encoding requires fewer bits and the extra bits are stored in the *bit reservoir*. Consequently, statistics of the *bit reservoir* of stego would be different. This idea was presented in [67]. Furthermore, recompression calibration was used to remove effect of audio contents from the extracted features. The system was tested on mp3 with different bit rates.

#### 4.1.5. Statistics of quantization step

Yan et al. observed that average value of the *quantization steps* between stegos and covers are the same but the difference between *quantization steps* of adjacent *granules* is increased [68]. According to the mp3 standard, both *granules* of the same frame use the same psychoacoustic model. Furthermore, it is logical to assume that consecutive *granules* have similar characteristics. This paper argued that embedding reduces such similarities. Assuming that the current *granule* is selected for embedding and its parity does not agree with the embedding bit, the algorithm increases the *quantization step* and adds the extra bits to the *bit reservoir*. If the next *granule* uses those extra bits, its *quantization step* is decreased. Therefore, the difference between *quantization steps* of two consecutive *granules* would increase.

### 4.2. Steganalysis of Ump3c

Ump3c is another steganography tool for mp3 that hides information in the LSB of *global gain* of each *granule*. Ump3c and Mp3stego have the same capacity, but unlike Mp3stego, Ump3c works directly on mp3 files.

Active steganalysis of Ump3c was conducted by Jin et al. [69]. Their method was a modified version of the regular-singular (RS) image steganalysis. In the RS steganalysis, an invertible flipping function is defined. Then effect of applying the flipping function on the noise of signal is evaluated. Based on this criterion three groups of regular, singular and neither are defined for the case of increase, decrease, or no change in the noise of signal. Typically covers have higher number of regular groups.

For steganalysis, *global gains* of mp3 were extracted and they were stored in the sequence $G = \{gg[k]\}$. Then, this sequence was segmented into groups of 4 samples with maximum overlap. If $gg_k$ denotes segment $k$, it is defined as:
$$gg_k = \{gg[k], gg[k+1], gg[k+2], gg[k+3]\} \quad (24)$$
Then, noise of each group was measured using equation (25).
$$\mathcal{N}(gg_k) = \sum_{i=0}^{2} |gg[k+i] - gg[k+i+1]| \quad (25)$$
Also, a set of flipping functions $F: x \leftrightarrow y$ were defined such that if their input is *x*, their output will be *y* and vice versa.
$$F_1: 0 \leftrightarrow 1, 2 \leftrightarrow 3, \dots, 254 \leftrightarrow 255 \quad (26)$$
$$F_{-1}: -1 \leftrightarrow 0, 1 \leftrightarrow 2, \dots, 255 \leftrightarrow 256 \quad (27)$$
Then, flipping function $F_1$ was applied on sequence *G* and its result was segmented and noise of each group was measured using equation (25). The same procedure was repeated for applying flipping function $F_{-1}$. Finally, the number of regular and singular groups from both flipping functions were used to estimate the length of hidden message.

### 4.3. Summary

Summary of reviewed papers are presented in Table 7.

### 5. Steganography and steganalysis of AAC

Advanced audio coding (AAC) is another popular compressed audio format which is used by many audio/video streaming services and websites. AAC is the successor of mp3 and it uses perceptual coding and entropy coding for achieving high compression rate while maintaining quality of signal. There are lots of similarities between the two, but ACC can achieve the same level of quality for 70% of bit rate of mp3 [70].

### 5.1. AAC steganography

AAC bit stream has different components and some of them have been used for steganographic purposes. They include Huffman table information, MDCT coefficients, and quantization parameters.

Possibility of changing Huffman coding section for embedding was pursued in [71]. Another possible place for data hiding is the sign bit of code words [72]. This work used an interesting technique for minimizing the distortion at the expense of reducing the capacity. If XOR of two consecutive sign bits did not agree with the message, sign bit of the smaller coefficient was changed [72]. LSB embedding in MDCT coefficients is another promising trend which was proposed in [73]. For this purpose, embedding was done during the encoding process. More specifically, embedding algorithm enforced usage of a smaller *scale factor* and exploited the extra bits for carrying the message.

Some other AAC embedding methods are direct adaptation from mp3 algorithms. Huffman tables of AAC can encode values of MDCT between 0 and 15 and therefore, frames with larger MDCTs use an especial symbol known as escape sequence. LSB of escape sequence in AAC was used for hiding information in [74]. This method is the counterpart of Linbit embedding in mp3 [57]. Bitrate steganography on AAC, hides message as the parity of the number of bits used for each frame [75] which is in parallel with mp3stego.

Table 7. Summary of mp3 steganalysis papers. Notations of the table are as follows SI: side information, BPB: bit per bit, R: maximum capacity ratio, -: not reported parameters, ×: non-applicable parameters

| Ref. | Target Method | Domain | Feature No. | Classifier | Train size | Database Spec. | | Performance | |
|---|---|---|---|---|---|---|---|---|---|
| | | | | | | Clean No. | Min Capacity | TNR | TPR |
| [7] | Mp3stego | SI | - | Threshold | - | - | 12.6% R | 84.4 | 79.3 |
| [63] | Mp3stego | MDCT | 214 | SVM | 75% | 1000 | 16% R | Ac. = 94.1 | |
| [64] | Mp3stego | MDCT | <200 | SVM | 60% | 5000 | 40% R | Ac. = 91.35 | |
| [65] | Mp3stego | MDCT | 64 kb:41 128 kb:115 192 kb:212 | SVM | - | 3000 | 10% R | 86.48 | 94.37 |
| [67] | Mp3stego | SI | 1 | SVM | 50% | 1200 | 0.001% BPB | 75 | 73.38 |
| [68] | Mp3stego | SI | 1 | Threshold | × | 1456 | 10% R | 80 | 96.72 |
| [69] | Ump3c | SI | × | × | × | 200 | 3.3% R | × | × |

## 5.2. AAC steganalysis

### 5.2.1. Calibrated Markov transition probability

Work of [76] showed that Huffman changing method [71], changes correlation between adjacent *scale factor* bands. Based on this observation, Markov transition probability between indexes of Huffman codebook of consecutive *scale factor* bands were used. The method only investigated correlation between tables 1 to 10, so steganalysis system was constructed with 100 features. Finally, potency of steganalysis features were improved by re-compression calibration. Simulation results showed that calibration can improves the results up to 10%.

### 5.2.2. Difference of inter and intra frame probabilities

Transitions in audio signals are very smooth, so adjacent samples are highly correlated in both time and frequency. This characteristic was exploited for steganalysis of Huffman table sign method [72]. This approach was proposed in [77] where statistical characteristics of inter-frame and intra-frame MDCT coefficients were used for steganalysis. The method constructed a rich model consisting of 1296 features and used ensemble classifier. To construct the rich model, based on their *block types*, AAC frames were divided into groups of short and long. Then, Markov transition probability and accumulative neighbouring joint density of first and second order derivative of inter and intra frames of each group were used as the final features.

## 6. Evaluation of Audio Steganalysis methods

### 6.1. Non-compressed methods

Reviewing previous works on audio steganalysis shows some shortcomings:

1- Most image steganalysis methods have been tested on BOSS or BOWS databases. Therefore, their reported results can be compared. On the other hand, in audio steganalysis such standard databases are not available and each work has used a different database. Evaluating performance of audio steganalysis methods on a single database makes a fair comparison between them possible, and alleviates this problem for future referencing.

2- Different audio steganography techniques have been proposed, most of which are non-LSB methods. Referring to table 4, it is evident that there is not a balance between the amount of works published on LSB and non-LSB methods. Therefore, non-LSB methods have not been investigated properly.

3- In practical situations the warden does not have any prior knowledge about the embedding algorithm. Thus, universal steganalysis resembles practical situations more closely. According to table 5, most of the previous works have only been investigated in the targeted scenario. Consequently, performance of many methods in universal scenario is not known.

4- Previous works have shown that complexity of signal plays an important role on the performance of steganalysis system [13]. But, most of existing works have not investigated this property.

In this section we try to fill these gaps. To that end, we used database of [20] which contained 4169 covers. Then, we embedded each cover with different methods and at different embedding capacities with random messages. LSB methods of Hide4pgp [35] and Steghide [9], non-LSB method of integer wavelet (I-wavelet) [46], and watermarking method of COX [31] were used for this purpose. In this manner our database had a total number of 54197 excerpts. We implemented some of the most recent and well-cited audio steganalysis methods and evaluated them in the both targeted and semi-blind universal paradigms. We used 10-fold cross validation and Matlab implementation of support vector machine (SVM) with "KernelScale" of Gaussian kernel, set to "auto". Furthermore, if a certain method had pre-processing (ex. feature normalization) or post-processing (feature selection) they were also implemented. Finally, complexity of all files were measured using equation (2) and then they were divided in three distinct sets. These regions were defined as follows.

*low complexity*: $C_{x[n]} \leq 0.06$ (28)
*medium complexity*: $0.06 < C_{x[n]} < 0.12$ (29)
*high complexity*: $0.12 \leq C_{x[n]}$ (30)

First, we evaluated each method in the targeted paradigm. Table 8 presents results of this analysis. The best results are shown in bold face letters.

Receiver operating characteristics (ROC) of different feature sets for detection of Steghide at capacity of 0.25 bit-per-symbol (BPS) for low and medium complexities are shown in figure 7.

Investigating results of table 8 and figure 7 shows that Cal-R-Mel has the best performance. Also, for most feature sets we see a negative correlation between complexity and performance. That is, signals with higher complexities are harder to detect. But at the same time, Markov feature does not follow this pattern. It is quite possible that if a better complexity measure is proposed, more meaningful results would be achieved.

Performance of different methods in the universal paradigm is compared. These results are shown in table 9 with the best results shown in the bold face letters.

Table 9. Comparison between different methods in the universal paradigm.

| Method | Ref. | Feature No | Complexity | Results | | |
|---|---|---|---|---|---|---|
| | | | | TPR | TNR | Ac. |
| D2-MFCC | [16] | 29 | low | 77.7 | 89.9 | 83.8 |
| | | | medium | 62 | 77.2 | 69.6 |
| | | | high | × | 59.5 | 50.4 |
| R-MFCC | [19] | 29 | low | 78.8 | 90.6 | 84.7 |
| | | | medium | 75.1 | 85.7 | 80.4 |
| | | | high | 61.8 | 76.1 | 69 |
| R-MFCC+HOS+GA | [20] | 7 | low | 92.6 | 93 | 92.8 |
| | | | medium | 88.8 | 88.8 | 88.8 |
| | | | high | 84.2 | 86.8 | 85.5 |
| Markov | [13] | 169 | low | 72.1 | 95.3 | 83.7 |
| | | | medium | 87.6 | 94.8 | 91.2 |
| | | | high | 94.4 | 95.5 | 94.9 |
| Cal-R-Mel | [28] | 15 | low | **99.6** | **98.8** | **99.2** |
| | | | medium | **99.7** | **98.7** | **99.2** |
| | | | high | **98.9** | **99.3** | **99.1** |

Table 8. Accuracy of different feature sets in targeted scenario. Capacity is expressed in terms of bit-per-symbol (BPS) and accuracies lower than random guess (50%) are marked by ×. The number of features in each method is mentioned after their name in the () and the reference number of each method is mentioned in [].

| Method | Capacity/ Param. | Complexity | D2-MFCC (29) [16] | R-MFCC (29) [19] | R-MFCC+ HOS+GA (7) [20] | Markov (169) [13] | Cal-R-Mel (15) [28] |
|---|---|---|---|---|---|---|---|
| **Hide4pgp** | C = 4 | low | 95.7 | 99.7 | 99.9 | **100** | **100** |
| | | medium | 83.6 | **100** | **100** | **100** | **100** |
| | | high | 63.4 | 98.6 | 99.9 | **100** | **100** |
| | C = 2 | low | 64.9 | 99.1 | 99.9 | **100** | **100** |
| | | medium | × | 97.1 | 99.5 | **100** | **100** |
| | | high | × | 91.2 | 98.7 | **100** | **100** |
| | C = 1 | low | × | 97.4 | 99.4 | 99.3 | **100** |
| | | medium | × | 90.8 | 97.4 | 99.9 | **100** |
| | | high | × | 74.8 | 95.1 | 99.7 | **99.9** |
| **Steghide** | C = 0.5 | low | × | 95 | 98.8 | 89.5 | **100** |
| | | medium | × | 86 | 95.7 | 95.3 | **100** |
| | | high | × | 68.5 | 91.5 | 98 | **100** |
| | C = 0.25 | low | × | 88.2 | 96.5 | 74.3 | **99.9** |
| | | medium | × | 76.5 | 90.2 | 88.6 | **99.6** |
| | | high | × | 59.1 | 85.6 | 95.2 | **100** |
| | C = 0.12 | low | × | 78.5 | 90.8 | 51 | **99.6** |
| | | medium | × | 66.2 | 83.8 | 73.1 | **99.5** |
| | | high | × | × | 79 | 87.3 | **99.5** |
| **I-wavelet** | C = 2 | low | 91.1 | 99.8 | 99.9 | **100** | **100** |
| | | medium | 72.5 | 99.6 | **100** | **100** | **100** |
| | | high | × | 97.5 | 99.9 | **100** | **100** |
| | C = 1 | low | 52.5 | 98.6 | 99.8 | 99.6 | **100** |
| | | medium | × | 95.8 | 99.2 | **100** | **100** |
| | | high | × | 85.6 | 97.5 | 99.9 | **100** |
| | C = 0.5 | low | × | 93.6 | 98 | 98.6 | **99.9** |
| | | medium | × | 83.8 | 94.2 | 99.7 | **99.8** |
| | | high | × | 62.1 | 89 | **99.7** | **99.7** |
| | C = 0.25 | low | × | 85.3 | 94.6 | 90.8 | **99.9** |
| | | medium | × | 71.4 | 86.2 | 96 | **99.5** |
| | | high | × | × | 77 | 98.7 | **99.6** |
| | C = 0.12 | low | × | 72.6 | 85.1 | 71.3 | **99.4** |
| | | medium | × | × | 70.1 | 89.3 | **98.9** |
| | | high | × | × | 56.5 | 96.3 | **98** |
| **COX** | α = 0.01 | low | 99.9 | 99.8 | 100 | **100** | 99.8 |
| | | medium | 97.7 | 99 | 99.7 | **100** | 99.4 |
| | | high | 87.6 | 98.8 | 99.7 | **100** | 99.7 |

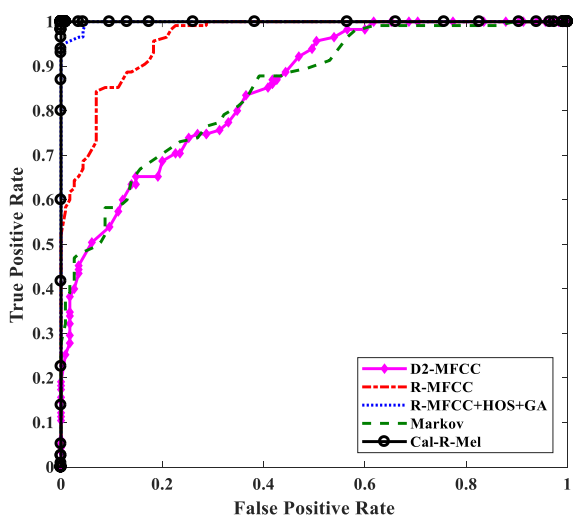
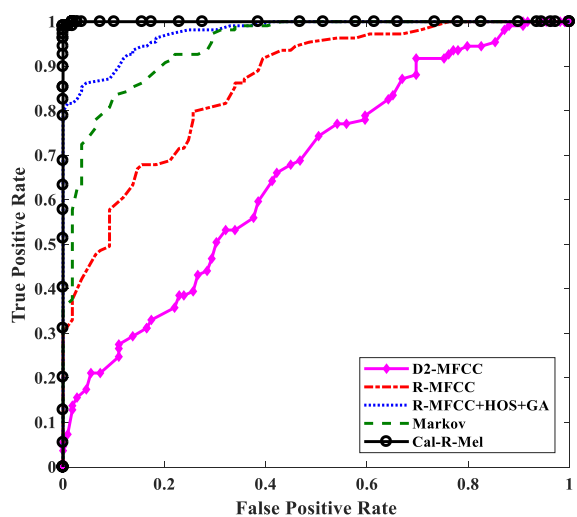

Figure 7. ROC of different feature sets for detection of Steghide at capacity of 0.25 BPS in the targeted paradigm
(A) low complexity   (B) medium complexity

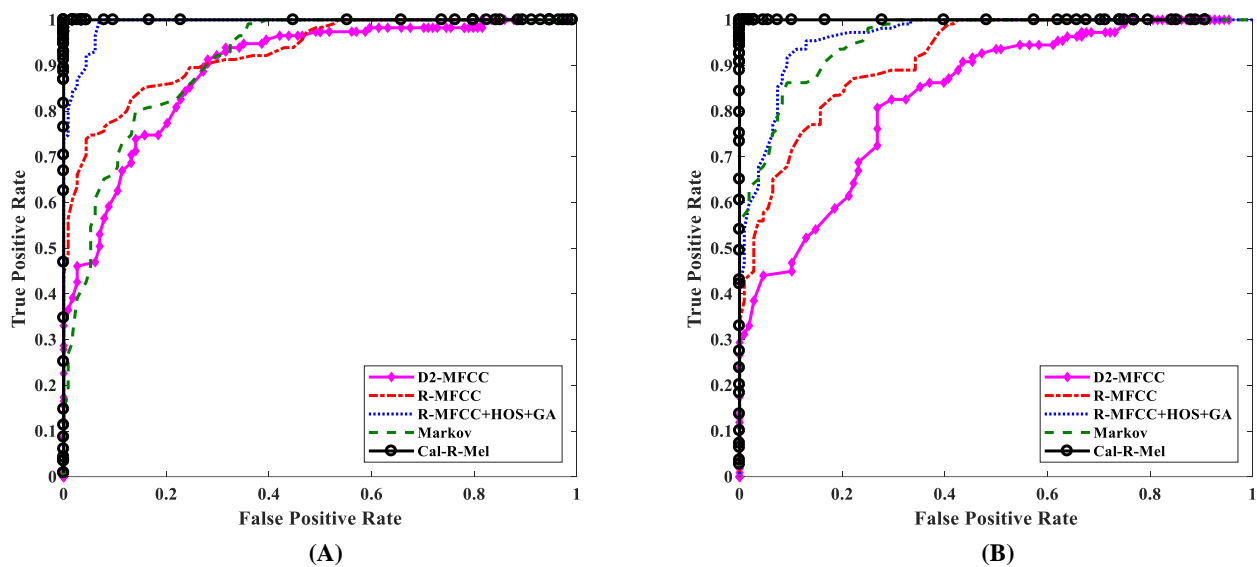

Figure 8. ROC of different feature sets in the universal paradigm   (A) low complexity   (B) medium complexity

ROC of different feature sets in the universal paradigm for low and medium complexities are shown in figure 8.

Comparing results of table 9 and figure 8 shows that Cal-R-Mel outperforms other methods by a large margin.

## 6.2. Mp3stego methods

Three different methods for steganalysis of Mp3stego were implemented and they were tested for different signal complexities. Results of these simulations are presented in table 10.

ROC of different feature sets for detection of Mp3stego at capacity of 12.5% for low and medium complexities are shown in figure 9.

Based on result of table 10 we can conclude that method of differential quantization, outperforms other steganalysis methods.

## 7. Future Works

This section discusses some directions that may be worth further investigation.

1- Steganalysis is implementation of a passive warden. On the other hand, active warden can measure robustness of steganography methods. This approach has not been addressed adequately in audio steganography. Investigating, robustness of existing methods to active warden and findings more robust methods can address this shortcoming.

2- Most of the reviewed papers have used the detection of watermarking systems (table 4) as an estimation of performance of steganalysis system on robust steganography methods. Comparing objectives of steganography and watermarking methods shows that undetectability is not the concern of watermarking. Therefore, reliable detection of watermarking systems cannot be interpreted as reliable detection of robust steganography methods. Detection of robust steganography methods should be analyzed separately.

3- Referring to table 4, it is evident that there is not a balance between the amount of works on LSB and non-LSB methods. Although, we tried to address this in section 6, yet non-LSB methods need more investigations.

4- According to table 5, most of the previous works have only investigated targeted steganalysis. We tried to address this by evaluating some methods under semi-blind scenario in section 6. Future works may focus on blind universal audio steganalysis.

5- A lot of powerful image steganalysis methods are based on calibration. A more thorough investigation of cover/stego estimation techniques in audio and proposing an efficient method would improve performance of steganalysis.

6- While there is a rich literature on steganalysis, cover space has not been studied properly. Specifically, the impact of cover contents (speech/music or genre of music), sampling frequency, quantization depth and etc. on steganalysis systems should be investigated.

7- The idea of signal complexity and its impact on steganalysis was proposed in [12]. Our analysis in section 6 showed that sometimes the existing metric does not work. Conducting a formal analysis on signal complexity and proposing a better measure for steganalysis applications would be fruitful for both steganography and steganalysis applications.

8- Most of current steganalysis methods process one frame of the signal and then use their moments for steganalysis. A framed based method is another approach that has not been considered yet. Such method may make a decision about every frame and then use an appropriate rule for making the final decision. Also, such methods would be very beneficial for steganography and its results can be used for better understanding the cover space and implementing adaptive steganography methods.

9- Previous works have shown the effect of audio contents on the result of steganalysis [11, 28]. Therefore, implementing a proper clustering method before steganalysis and investigating its effect seems to be fruitful.

Table 10. Accuracy of different feature sets. Capacity is expressed as percentage of maximum embedding capacity. The number of features in each method is mentioned after their name in the () and the reference number of each method is mentioned in [].

| Capacity | Complexity | MDCT (48) [64] | | | Differential Quantization (1) [68] | | | DAMDCT (34) [66] | | |
|---|---|---|---|---|---|---|---|---|---|---|
| | | TPR | TNR | Ac. | TPR | TNR | Ac. | TPR | TNR | Ac. |
| 100 | low | 95 | 95 | 95 | **96.6** | 94.6 | **96.6** | 92.9 | **95.5** | 94.2 |
| | medium | 94.4 | 94.9 | 94.6 | **97.4** | **95.7** | **97.4** | 92.6 | 94.3 | 93.5 |
| | high | 95.6 | 94.9 | 95.3 | **96.7** | 94.9 | **96.7** | 91.8 | **95.4** | 93.6 |
| 50 | low | 90.8 | 90.8 | 90.8 | **95.5** | 93 | **95.5** | 93.8 | **95.3** | 94.6 |
| | medium | 91 | 90 | 90.5 | **95.7** | 93.6 | **95.7** | 91.6 | **94** | 92.8 |
| | high | 91.7 | 90.4 | 91.1 | **94.5** | 92.6 | **94.5** | 92.1 | **95.3** | 93.7 |
| 25 | low | 88.4 | 87.1 | 87.8 | **94.2** | 92.3 | **94.2** | 93.4 | **94.3** | 93.9 |
| | medium | 88.6 | 86.8 | 87.7 | **95.4** | 92.6 | **95.4** | 91.4 | **94.5** | 92.9 |
| | high | 89.2 | 87.1 | 88.1 | **91.7** | 87.7 | **91.7** | 91.5 | **94.4** | 93 |
| 12.5 | low | 87.8 | 86.8 | 87.3 | **94.3** | 91.3 | **94.3** | 93.2 | **94.5** | 93.9 |
| | medium | 86.6 | 86.6 | 86.6 | **95.3** | 92.5 | **95.3** | 91.2 | **94.4** | 92.8 |
| | high | 89.3 | 87.8 | 88.6 | **92.2** | 88 | 92.2 | 92.6 | **93.9** | 93.3 |

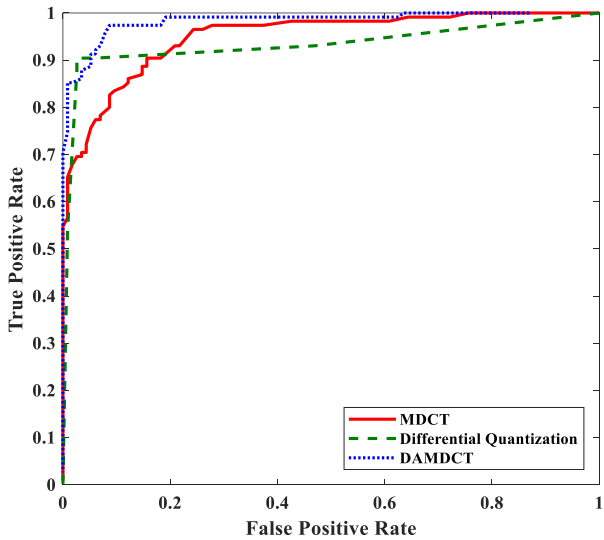
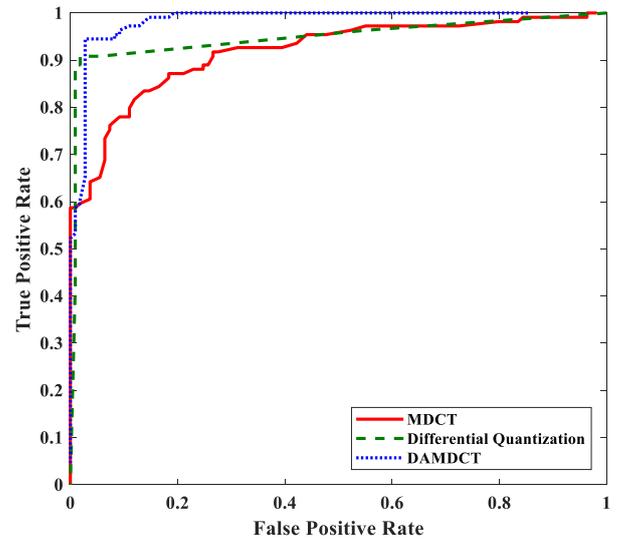

Figure 9. ROC of different feature sets for detection of mp3stego at capacity of 12.5%
(A) low complexity   (B) medium complexity

10- Current literature has only considered steganalysis of mono signals. Investigating the correlation between different channels of stereo signals may improve performance of steganalysis.
11- Referring to table 5, most methods have used SVM for classification. On the other hand, recent image steganalysis methods suggest that better performances may be achieved if rich models with ensemble classifiers are employed. Furthermore, recent trends in pattern recognition tasks have turned considerably in favor of deep learning techniques. But, such approaches have not been investigated for audio steganalysis.
12- The main focus of mp3 steganalysis methods has been on Mp3stego. Consequently, most of mp3 methods have not been investigated at all. Steganalysis of those methods should be addressed.
13- Different encoders for mp3 are available and they exhibit quite different characteristics [78, 79]. Studying the effect of those differences on steganalysis systems seems another direction that needs more investigation.
14- Due to wide spread usage of AAC in many audio/video streaming services it could be one of the best covers for steganography. Unfortunately, neither steganography nor steganalysis of AAC has found adequate attentions.

## 8. Conclusion

This work conducted a comprehensive review of audio steganalysis literature and classified them into different categories. Reviewing the audio steganalysis literature showed that their main contributions had been their feature extractions. Therefore, we specifically paid more attention to that part. Considering the type of cover signal, existing methods were broadly divided into non-compressed (wave) and compressed (mp3 and AAC) methods. Furthermore, considering the absence/presence of knowledge about the embedding algorithm, the systems were divided into two sub-groups of universal and targeted methods. For a better comparison between different works, each subsection of the paper was concluded with a summary of the relevant papers. Also, to conduct a fair comparison between different methods, some of them were implemented and were tested on the same database, on both LSB and non-LSB steganography methods, and in both targeted and universal scenarios. In the end, some future directions for audio steganalysis were discussed.

## Acknowledgments

The authors would like to thank the anonymous reviewers for their constructive comments and valuable contributions.